\documentclass[preprint,pra,nofootinbib]{revtex4}
\usepackage{mathrsfs}
\usepackage{graphicx}
\usepackage{epsfig}
\usepackage{dcolumn}% Align table columns on decimal point
\usepackage{bm}% bold math
\usepackage{amsmath,amssymb,amsthm}
\usepackage[colorlinks=true,linkcolor=blue]{hyperref}
\usepackage{subfigure}
\usepackage{booktabs}
\usepackage[mathscr]{euscript}
\usepackage{ulem}

\newcommand{\bea}{\begin{eqnarray}}
\newcommand{\eea}{\end{eqnarray}}
\newcommand{\beq}{\begin{equation}}
\newcommand{\eeq}{\end{equation}}

\def\/{\over}

\begin{document}
\title{Alice-Bob Peakon Systems}
\author{S. Y. Lou$^{1,2}$ and Zhijun Qiao$^3$}
\affiliation{%$^1$ University of Texas ...\\
$^{1}$Center for Nonlinear Science and Department of Physics, Ningbo University, Ningbo, 315211, China\\
$^2$Shanghai Key Laboratory of Trustworthy Computing, East China Normal University, Shanghai 200062, China\\
$^3$School of Mathematical and Statistical Science, University of Texas -- Rio Grande Valley, Edinburg, TX 78539, USA}

\begin{abstract}

In this letter, we study the Alice-Bob peakon system generated from an integrable peakon system through using the strategy of the so-called Alice-Bob non-local KdV approach \cite{LH1}. Nonlocal integrable peakon equations are obtained and shown to have peakon solutions.

\end{abstract}

\maketitle

\section{Introduction}

Recently, nonlocal integrable systems have attracted much attention since the pioneer work about nonlocal nonlinear Schr\"odinger equation \cite{AM1}. Lou and Huang applied the nonlocal approach to the KdV equation and proposed the so-called Alice-Bob Physics to generate coherent solutions of nonlocal KdV systems \cite{LH1}.
Before discussing AB peakon systems, let us first recall the basic procedure for regular AB integrable systems.

In the study of classical physics theory,   most feasible models are locally established around a single space-time point, say, $\{x,\ t\}$.
To investigate some related physical phenomena in two or more places, one has to consider some kinds of new models.
In papers \cite{8,LH1}, the authors proposed Alice-Bob (AB) models to study two-place physical problems. Alice-Bob physics makes sense if the physics is related to
two entangled events occurred in two places $\{x,\ t\}$ and $\{x',\ t'\}$. The event at
$\{x,\ t\}$ is called Alice event (AE) (denoted by $A(x,\ t)$)
and the event at $\{x',\ t'\}$ is called Bob event (BE) (denoted by $B(x',\ t')$).
The events AE and BE are called correlated/entangled if AE happens,  BE can be determined immediately by the correlation condition
\begin{equation}
B(x',\ t')=f(A)=A^f=\hat{f}A,
\end{equation}
where $\hat{f}$ stands for a suitable operator, which may be different at different events.
In general, $\{x',\ t'\}$ is not a neighbour to $\{x,\ t\}$. Thus, the intrinsic two-place models, the Alice-Bob systems (ABSs), are nonlocal.
Some special types of two-place nonlocal models have been proposed. For example, the following nonlocal nonlinear Schr\"odinger (NLS) equation,
$$
iA_t+A_{xx}\pm A^2B=0,\  B=\hat{f}A=\hat{P}\hat{C}A=A^*(-x,t), $$
was firstly proposed by  Ablowitz and Musslimani\cite{AM1}.
This type of NLS systems has strong relationships to the significant PT symmetric  Schr\"dinger equations\cite{AMa}. The operators
$\hat{P}$ and $\hat{C}$ are the usual parity and charge conjugation. Afterwards,
   other nonlocal systems, including the coupled nonlocal NLS systems \cite{DNLS}, the nonlocal modified KdV systems \cite{MKdV,MKdVa}, the discrete nonlocal NLS system \cite{dNLS}, and the nonlocal Davey-Stewartson systems \cite{DS,DSa,DSb}, were investigated as well.

In this letter, we shall utilize the Alice-Bob approach to study integrable peakon equations. For our convenience, the peakon system based on the Alice-Bob (AB) approach is called the Alice-Bob peakon (ABP) system. We will take the integrable peakon systems studied in \cite{XQZ1} as the ABP examples.

\section{An integrable Alice-Bob peakon system with Lax pair and infinitely many conservation laws}

Let us consider the following model:
\begin{eqnarray}
&&m_t=(mH)_x+mH-\frac12m(A-A_x)(\bar{A}+\bar{A}_x),\ \quad \bar{A}=A(-x+x_0,\ -t+t_0), \label{ABP}\\
&& m=A-A_{xx},
\end{eqnarray}
where $H$ is an arbitrary shifted parity ($\hat{P}_s$) and delayed time reversal ($\hat{T}_d$) invariant functional
\begin{eqnarray}
\hat{P}_s\hat{T}_d H=\bar{H}=H,
\end{eqnarray}
and the definitions of $\hat{P}_s$ and $\hat{T}_d$ read $ \hat{P}_s x=-x+x_0,\quad \hat{T}_d t=-t+t_0. $
Actually, equation (\ref{ABP}) is generated through the first equation of the two-component system (7) proposed in the paper \cite{XQZ1} via $u=A$  and $v=\bar{A}$.

Through a lengthy computation, one may have a Lax pair for equation (\ref{ABP})
\begin{eqnarray}
&& \phi_x=U\phi,\quad U
\left(\begin{array}{cc}
-1 & \lambda m\\
-\lambda\bar{m} & 1
\end{array}\right),\quad \bar{m}=\bar{A}-\bar{A}_{xx},\\
&& \phi_t=V\phi,\quad V=
\frac12\left(\begin{array}{cc}
\frac12(A_x-A)(\bar{A}+\bar{A}_x)-\lambda^{-2} & \lambda^{-1}(A-A_x)+\lambda m H\\
-\lambda^{-1}(\bar{A}+\bar{A}_x)-\lambda \bar{m} H & \lambda^{-2}+\frac12(A-A_x)(\bar{A}+\bar{A}_x)
\end{array}\right).
\end{eqnarray}

Following the typical procedure starting from Lax pair, we may obtain the following conservation laws for equation (\ref{ABP}):
\begin{eqnarray}
&&\rho_{jt}=J_{jx},\ j=0,\ 1,\ 2,\
\ldots,\ \infty,\\
&& \rho_j=m\omega_j,\ j=0,\ 1,\ 2,\ \ldots, \\
&& J_j=(A-A_x)\omega_{j-2}+H\rho_j,\ j=2,\ 3,\ \ldots, \\
&& J_0=H\rho_0,\ J_1=H\rho_1-\frac12(A-A_x)(\bar{A}+\bar{A}_x),\\
&& \omega_{j+1}=\frac1{m\omega_0}\left(\omega_j-\omega_{jx}-\frac{m}2
\sum_{k=1}^j\omega_k\omega_{j+1-k} \right),\ j=1,\ 2,\ \ldots,\\
&& \omega_0=\sqrt{-\bar{m}m^-1},\quad \omega_1=\frac{m\bar{m}_x-\bar{m}m_x-2m\bar{m}}{2m^2\bar{m}}.
\end{eqnarray}
It is noticed that the conserved densities $\rho_j=m\omega_j,\ j=0,\ 1,\ \ldots,$ are not explicitly  dependent on $H$.

\section{Peakon solutions for special ABP systems}

Let us now solve concrete peakon solutions from equation (\ref{ABP}) for some special functions $H$ listed in the following examples.

\bf Example 1. \rm Taking $H=0$ sends equation (\ref{ABP}) to the following system
\begin{eqnarray}
\left\{\begin{array}{l}
m_t=-\frac12 m(A-A_x)(\bar{A}+\bar{A}_x),\label{Ex1}
 \\
m=A-A_{xx},\ \bar{A}\equiv A(-x+x_0,\ -t+t_0).
\end{array} \right.\nonumber
\end{eqnarray}

This system has only one-peakon solutions:
\begin{equation}
A=c \exp\left[-\frac13 c^2\left(t-\frac{t_0}2 \right)-\left|x-\frac{x_0}{2}\right|\right],\label{p1}
\end{equation}
which are non-traveling solitary waves with a fast decayed standing peak and shown in Fig. 1.

\input epsf
\begin{figure}[tbh]
\centerline{\epsfxsize=9.0cm\epsfysize=13cm\epsfbox{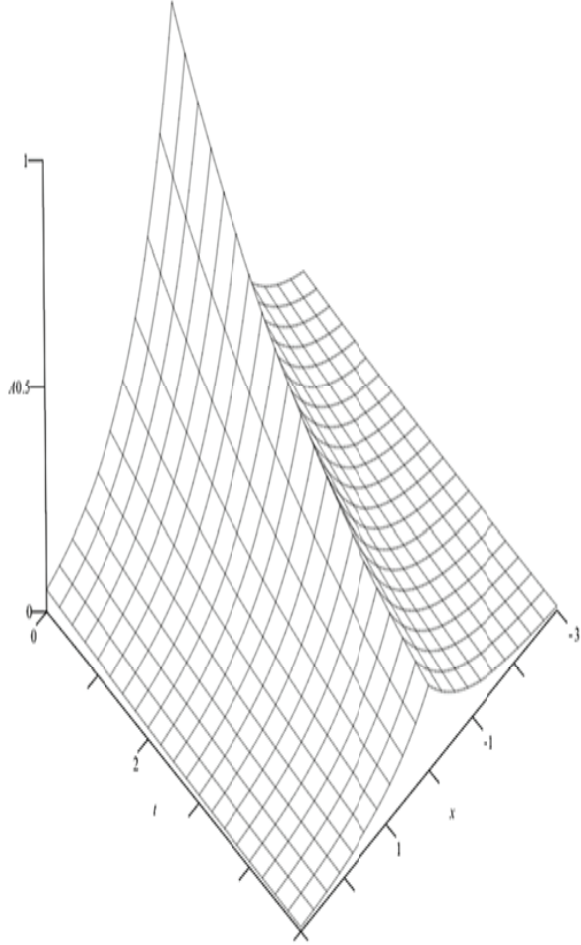}}
\caption{Figure 1. The fast decayed non-traveling peakon solution \eqref{p1} for the first special ABP system \eqref{Ex1} with the peakon parameters $ x_0=t_0=0$ and $c=1$.}
\label{fig1}
\end{figure}

No multi-peakon solution is found for this special example though it is integrable with Lax pair and infinitely many conservation laws.

\bf Example 2. \rm Selecting
$H=\frac12
(A\bar{A}-A_x\bar{A}_x)$ leads equation (\ref{ABP}) to
\begin{eqnarray}
\left\{\begin{array}{l}
m_t=\frac12\big[m(A\bar{A}-A_x\bar{A}_x)\big]_x-\frac12 m(A-A_x)(\bar{A}+\bar{A}_x),\label{Ex2}
 \\
m=A-A_{xx},\ \bar{A}\equiv A(-x+x_0,\ -t+t_0).
\end{array} \right.
\end{eqnarray}

This equation possesses the following  one peakon solutions
\begin{equation}
A=c \exp\left[-\left|\left(x-\frac{x_0}{2}\right)+\frac13 c^2\left(t-\frac{t_0}2 \right)\right|\right].
\label{p2}
\end{equation}
as well as $N$-peakon dynamical system
\begin{eqnarray}
&&A=\sum_{j=1}^N p_j \exp\left(-\left|x-\frac{x_0}{2}-q_j\right|\right),\label{peakn1}
\end{eqnarray}
\begin{eqnarray}
&&p_{jt}=\frac{1}{2}p_j\sum_{i,k=1}^Np_i\bar{p}_k[sgn(q_j-q_k)-sgn(q_j-q_i)]
e^{-|q_j-q_k|-|q_j-q_i|},\\
&&q_{jt}=\frac{1}{6}p_j\bar{p}_j
-\frac12\sum_{i,k=1}^Np_i\bar{p}_k[1-sgn(q_j-q_k)sgn(q_j-q_i)]
e^{-|q_j-q_k|-|q_j-q_i|}.
\end{eqnarray}

Though the integrability problem of the above $N$-peakon system is open, we do get the following explicit 2-peakon solutions
\begin{eqnarray}
&&p_{1t}=\frac{1}{2}p_1(p_1\bar{p}_2-p_2\bar{p}_1)sgn(q_1-q_2)
e^{-|q_1-q_2|}),\\
&&p_{2t}=\frac{1}{2}p_2(p_2\bar{p}_1-p_1\bar{p}_2)sgn(q_2-q_1)
e^{-|q_2-q_1|}),\\
&&q_{1t}=-\frac13 p_1\bar{p}_1-\frac12(p_1\bar{p}_2+p_2\bar{p}_1)e^{-|q_1-q_2|},\\
&&q_{2t}=-\frac13 p_2\bar{p}_2-\frac12(p_1\bar{p}_2+p_2\bar{p}_1)e^{-|q_2-q_1|}.
\end{eqnarray}
Therefore,
\begin{eqnarray}
&&q_1=\frac{3p_1p_2 sgn(2t-t_0)}{|p_1^2-p_2^2|}\left[e^{-\frac16 |(p_1^2-p_2^2)(2t-t_0)|}-1\right]-\frac{p_1^2}6(2t-t_0),
\nonumber\\
&&q_2=\frac{3p_1p_2 sgn(2t-t_0)}{|p_1^2-p_2^2|}\left[e^{-\frac16 |(p_1^2-p_2^2)(2t-t_0)|}-1\right]-\frac{p_2^2}6(2t-t_0),
\label{q1q2}
\end{eqnarray}
where $p_1$ and $p_2$ are two arbitrary constants.

Fig.2 exhibits the single steady traveling peakon solution expressed by
\eqref{p2}.
\input epsf
\begin{figure}[tbh]
\centerline{\epsfxsize=9.0cm\epsfysize=13cm\epsfbox{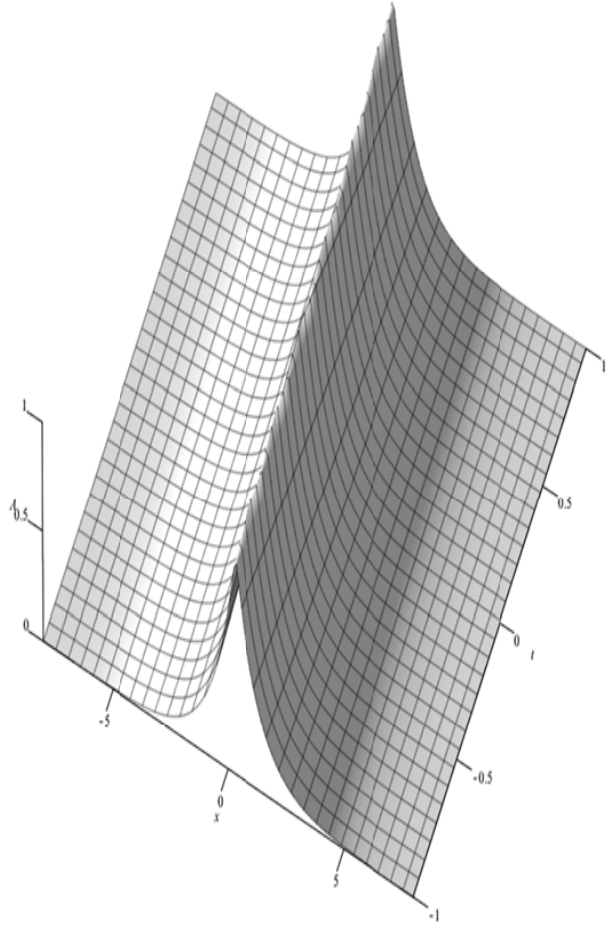}}
\caption{Figure 2. The single steady traveling peakon solution \eqref{p2} for the second special ABP system \eqref{Ex2} with the peakon parameters $ x_0=t_0=0$ and $c=1$.}
\label{fig2}
\end{figure}

Fig.3 shows the interactional behaviour for two-peakons given by \eqref{peakn1} with $N=2$ and \eqref{q1q2}.
\input epsf
\begin{figure}[tbh]
\centerline{\epsfxsize=9.0cm\epsfysize=13cm\epsfbox{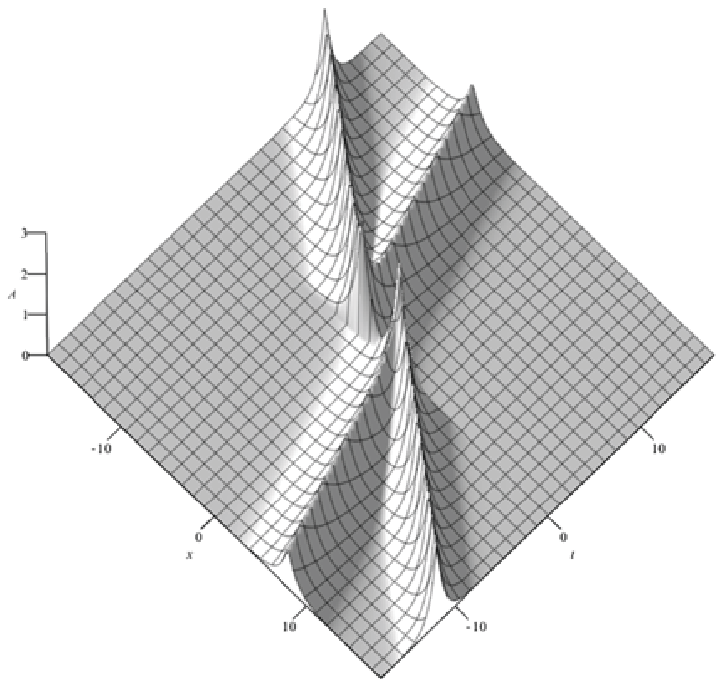}}
\caption{Figure 3. The two peakon interactional solution \eqref{peakn1} for the second special ABP system \eqref{Ex2}. The corresponding parameters are chosen as $N=2,\ x_0=t_0=0,\ p_1=1$ and $p_2=2$.}
\label{fig3}
\end{figure}

\bf Example 3. Choosing \rm
$H=\frac12
(A\bar{A}_x-A_x\bar{A})$ casts equation (\ref{ABP}) to the following system
\begin{eqnarray}
\left\{\begin{array}{l}
m_t=\frac12\big[m(A\bar{A}_x-A_x\bar{A})\big]_x-\frac12m\big[A\bar{A}-A_x\bar{A}_x\big],\
 \\
m=A-A_{xx},\ \bar{A}\equiv A(-x+x_0,\ -t+t_0),
\end{array} \right.
\end{eqnarray}
which has fast decayed one-peakon solutions as follows
\begin{equation}
A=c \exp\left[-\frac13 c^2\left(t-\frac{t_0}2 \right)-\left|x-\frac{x_0}{2}\right|\right].
\end{equation}
This single peakon possesses the same form as the one in Example 1.

Similar to the example one we have not yet found two-peakon solutions though the model is also integrable.

\bf Example 4. Let \rm
$H=\frac12
(A-A_x)(\bar{A}+\bar{A}_x)$. Then we obtain the following system
\begin{eqnarray}
\left\{\begin{array}{l}
m_t=\frac12\big[m(A-A_x)(\bar{A}+\bar{A}_x)\big]_x,\
 \\
m=A-A_{xx},\ \bar{A}\equiv A(-x+x_0,\ -t+t_0),
\end{array} \right.
\end{eqnarray}
which admits the following one-peakon solutions  %has the form
\begin{equation}
A=c \exp\left[-\left|\left(x-\frac{x_0}{2}\right)+\frac13 c^2\left(t-\frac{t_0}2 \right)\right|\right].
\end{equation}

Furthermore, we can obtain $N$-peakon dynamical system %, we have
\begin{eqnarray}
&&A=\sum_{j=1}^N p_j \exp\left(-\left|x-\frac{x_0}2-q_j\right|\right)
\end{eqnarray}
\begin{eqnarray}
&&p_{jt}=0,\\
&&q_{jt}=\frac{1}{6}p_j\bar{p}_j
-\frac12\sum_{i,k=1}^Np_i\bar{p}_k[sgn(q_j-q_k)-1][sgn(q_j-q_i)]+1]
e^{-|q_j-q_k|-|q_j-q_i|}.
\end{eqnarray}

In particular, two peakon solutions take on the following form
\begin{eqnarray}
&&q_1=\frac16 p_1^2(2t-t_0)
+\frac{3p_1p_2 sgn(2t-t_0)}
{|p_1^2-p_2^2|}\left(e^{-\frac16|(p_1^2-p_2^2)(2t-t_0)|}-1\right),\\
&&q_2=\frac16 p_2^2(2t-t_0)
+\frac{3p_1p_2 sgn(2t-t_0)}
{|p_1^2-p_2^2|}\left(e^{-\frac16|(p_1^2-p_2^2)(2t-t_0)|}-1\right)
\end{eqnarray}
where $p_1$ and $p_2$ are two arbitrary constants. In this case, both the single peakon and two peakon forms are the same as those in Example 2.

\section{Summary and discussion}
In this paper, we give a quite general integrable Alice-Bob peakon system \eqref{ABP} with an arbitrary $\hat{P}_s\hat{T}_d$ invariant
functional. The existence of multi-peakon solutions are discussed for some special cases through selecting different invariant functionals $\hat{P}_s\hat{T}_d$.
The Lax pair of the ABP system \eqref{ABP} and infinitely many conservation laws are found. Starting from the Lax pair and the conservation laws, one may readily set up other integrable properties such as the infinitely many symmetries, bi-Hamiltonian structures, recursion operator, and inverse scattering transformation. %can be readily obtained.
But, we do not discuss those topics here in our paper, instead, we focus on peakon solutions for the ABP system.
It already reveals from Examples 1 and 3 that for the ABP systems if the single peakon is non-travelling (standing) and fast decayed, then there may be no two-peakon interactional solutions.
If the single peakon is non-standing (travelling) and not decayed like Examples 2 and 4, then there may exist multi-peakon solutions.
Another amazing fact is %also found
that for different ABP systems like Examples 2 and 4, their multi-peakon solutions might be completely same.

There are some intrinsic difference between  usual peakon systems and  ABP systems. The main difference is that usual peakon systems are local while the ABP systems are nonlocal. The nonlocal property is introduced because the model includs two far-away correlated events, namely, AE and BE. Due to the intrusion of two far-away events in one model, the invariant property in both space and time translations is broken.
 This kind of symmetry-breaking property probably destroys the existence of multiple decayed standing peakons.

The study of regular peakon systems is one of the hot topics in mathematical physics and many different peakon systems were found in the literature.
For every peakon system, there may exist different versions of integrable ABP systems. Here in the following we just list some of them we developed, but leave the details for our near future investigations.

\bf Alice-Bob Cammasa-Holm (ABCH). \rm Based on the well-known Camassa-Holm (CH) equation \cite{CH1}, one of the integrable ABCH systems can be written as
$$(1-\partial_x^2)A_t=(A+B)(3A-A_{xx})_x-2(A+B)_xA_{xx}+H $$
where $B=\hat{P}_s\hat{T}_dA$ and $H$ is an arbitrary $\hat{P}_s\hat{T}_d$ invariant functional.

\bf Alice-Bob DP (ABDP). \rm Similar to the ABCH, adopting the well-known Degasperis-Procesi (DP) equation \cite{DP1} yields the integrable ABDP system in the following form
$$(1-\partial_x^2)A_t=(A+B)(4A-A_{xx})_x-3(A+B)_xA_{xx}+H $$
with an arbitrary $\hat{P}_s\hat{T}_d$ invariant  functional $H$ and $B=\hat{P}_s\hat{T}_dA$.

%%%%%%%%%%%%%%%% add b-family, check if this is correct?
\bf Alice-Bob $b$-family (ABbf). \rm Similar to the ABCH and ABDP and considering the $b$-family equation \cite{HS1}, one may generate the ABbf system in the following form
$$(1-\partial_x^2)A_t=(A+B)((b+1)A-A_{xx})_x-b(A+B)_xA_{xx}+H $$
with an arbitrary $\hat{P}_s\hat{T}_d$ invariant  functional $H$ and $B=\hat{P}_s\hat{T}_dA$.
%%%%%%%%%%%%%%%

\bf Alice-Bob Novikov (ABN). \rm For the Novikov equation \cite{Nov1}, there exists the following ABN system
$$(1-\partial_x^2)A_t=(A+B)^2(A_{xx}-4A)_x+\frac32(A+B)^2_xA_{xx}+H $$
with $\hat{P}_s\hat{T}_d$ invariant  functional $H$ and $B=\hat{P}_s\hat{T}_dA$.

\bf Alice-Bob FORQ (ABFORQ). \rm For the FORQ system \cite{F1,OR1,QJMP1}, we have an integrable ABFORQ extension as follows
$$(1-\partial_x^2)A_t=\left\{[(A+B)^2-(A+B)^2_x](A_{xx}-A)\right\}_x+H $$
where $H$ is $\hat{P}_s\hat{T}_d$ invariant and $B=\hat{P}_s\hat{T}_dA$.

\section*{Acknowledgements}
The author are in debt to the helpful discussions with Professors X. B. Hu and Q. P.  Liu.
The work was sponsored by the Global Change Research
Program of China (No.2015CB953904), the National Natural Science Foundations of China (Nos. 11435005), Shanghai Knowledge Service Platform for Trustworthy Internet of Things (No. ZF1213) and K. C. Wong Magna Fund at Ningbo University. The author (Qiao) thanks the UTRGV President't Endowed Professorship and the UTRGV
College of Science seed grant for their partial supports.


\begin{thebibliography}{00}
\bibitem{AM1}Ablowitz, M. J. and  Musslimani, Z. H.  Integrable nonlocal nonlinear Schr\"odinger equation. \it Phys. Rev. Lett. {\bf 110} \rm (2013) \rm, 064105.% (2013).
\bibitem{dNLS}Ablowitz, M. J. and Musslimani, Z. H.  Integrable discrete PT symmetric model, {\it Phys. Rev. E.} {\bf 90} (2014) \rm, 032912.% (2014).
\bibitem{MKdV}Ablowitz, M. J. and Musslimani, Z. H. Inverse scattering transform for the integrable nonlocal nonlinear Schrödinger equation. \it Nonlinearity, {\bf 29} \rm (2016) \rm, 915-946.% (2016).
\bibitem{CH1}Camassa, R. and Holm, D.D. An integrable shallow water equation with peaked solitons. {\it Phys. Rev. Lett.}, {\bf 71} (1993) 1661-1664.
\bibitem{DP1}Degasperis, A. and Procesi, M. Asymptotic integrability, in A. Degasperis and G. Gaeta (eds), {\it Symmetry and Perturbation Theory,} World Scientific, 1999, pp. 23-37.

\bibitem{DS} Dimakos, M. and Fokas, A. A.  Davey-Stewartson type equations in 4+2 and 3+1 possessing soliton solutions. \it J. Math. Phys. {\bf 54} \rm 2013 \rm, 081504.% (2013).
\bibitem{DSa} Fokas, A. S. Integrable Nonlinear Evolution Partial Differential Equations in
4+2 and 3+1 Dimensions. \it Phys. Rev. Lett. {\bf 96} \rm (2006) \rm, 190201.
\bibitem{F1}Fokas, A. S. On a class of physically important integrable equations, {\it Physica D} {\bf 87} (1995), 145-154.
%\bibitem{FoFu}Fokas A S and  Fuchssteiner B, Symplectic structures, their B\"acklund transformations and hereditary symmetries, {\it Phys. D}, {\bf 4}(1981), 47-66.
%\bibitem{Fu1}Fuchssteiner, B. Some tricks from the symmetry-toolbox for nonlinear equations: Generalizations of the Camassa-Holm equation,  {\it Physica D} {\bf 95}(1996), 229-243.

\bibitem{DSb}Fokas, A. S. Integrable multidimensional versions of the nonlocal nonlinear Schr\"odinger equation. \it Nonlinearity. {\bf 29} \rm (2016) \rm, 319-324.


\bibitem{HS1}Holm, D. D. and Staley, M. F. Nonlinear balance and
exchange of stability in dynamics of solitons, peakons, ramps/cliffs
and leftons in a $1+1$ nonlinear evolutionary PDE, {\it Phys. Lett.
 A}, {\bf 308} (2003), 437-444.

\bibitem{MKdVa}Ji, J. L. and Zhu, Z. N. Soliton solutions of an integrable nonlocal modified Korteweg-de Vries equation through inverse scattering transform.
https://arxiv.org/pdf/1603.03994.pdf, March 2016.
\bibitem{8}Lou, S. Y. Alice-Bob systems, $P_{s}-T_{d}-C$ principles and multi-soliton solutions. arXiv: 1603. 03975v2. nlin. SI, March 2016.
\bibitem{LH1}Lou, S. Y. and Huang, F  Alice-Bob Physics, coherent solutions of nonlocal KdV systems, Scientific Reports 7, Article number: {\bf 869} (2017); arXiv: 1606. 03154v1. nlin. SI, June 2016.
\bibitem{AMa}Musslimani, Z. H., Makris, K. G., El-Ganainy, R. and Christodoulides, D. N. \it Phys. Rev. Lett. {\bf 100} \rm (2008) \rm, 030402.
\bibitem{Nov1}Novikov, V. S. Generalisations of the Camassa-Holm equation, arXiv:0905.2219v1, May 2009.
\bibitem{OR1}Olver, P. and  Rosenau, P. Tri-Hamiltonian duality between
 solitons and solitary-wave solutions having compact support,
{\it Phys. Rev. E}, {\bf 53} (1996) 1900-1910.
 \bibitem{QJMP1}Qiao, Z. J.  A new integrable integrable equation with
 cuspons  and
M/W-shape peak solitons, {\it J. Math. Phys.}, {\bf 47} (2006),
112701-08; {\bf 48} (2007), 082701-20.

\bibitem{DNLS}Song, C. Q., Xiao, D. M. and Zhu, Z. N.  A General Integrable Nonlocal Coupled Nonlinear Schrödinger Equation. \rm https://arxiv.org/pdf/1505.05311.pdf, May 2015.
\bibitem{XQZ1}Xia, B. Q., Qiao, Z. J. and Zhou, R. G. A Synthetical Two-Component Model with Peakon Solutions. \it Studies in Applied Mathematics, {\bf 135} \rm (2015), \rm 248–276.

\end{thebibliography}
\end{document}